\documentclass{llncs}
\usepackage{fontspec}
\usepackage{graphicx}
\usepackage{svg}
\usepackage{subcaption}
\usepackage{placeins}
\usepackage[utf8]{inputenc}
\usepackage[T1]{fontenc}
\usepackage{amsmath}
\usepackage{amssymb}
\usepackage[hidelinks]{hyperref}
\usepackage{tikz}
\usepackage{xcolor}
\usepackage{fontspec}
\usepackage{color,soul}
\usetikzlibrary{arrows.meta, positioning}
\usepackage{url}

\begin{document}
\mainmatter              % start of a contribution
\title{Multimodal Smart Glove for Sign Language Recognition Using Deep Learning}
\titlerunning{}  % abbreviated title (for running head)
\author{
Anh Thu Nguyen Ngoc\inst{1} \and
Tam Phong Truong\inst{1} \and
Thai Anh Nguyen Duong\inst{1} \and
Vu Linh Nguyen\inst{2} \and
Manh Duong Phung\inst{2}
}

\authorrunning{Anh Thu Nguyen Ngoc et al.}

\tocauthor{Anh Thu Nguyen Ngoc, Tam Phong Truong, Thai Anh Nguyen Duong, Vu Linh Nguyen, Manh Duong Phung}

\institute{
Fulbright University Vietnam, Ho Chi Minh City, Vietnam \and
College of Engineering and Computer Science and Smart Green Transformation Center (GREEN-X), VinUniversity, Hanoi, Vietnam
}

\maketitle{}              % typeset the title of the contribution

\begin{abstract}

Sign language recognition technologies can improve communication between deaf individuals and the broader community, but many existing systems face challenges in real-world deployment. This paper presents a deployable smart glove system for sign language recognition that integrates wearable sensing and deep learning. The glove incorporates flex sensors and an inertial measurement unit (IMU) to capture finger articulation and hand motion, while facial cues are obtained through a camera. Sensor data are transmitted via an ESP32-C6 microcontroller and processed using a long short-term memory (LSTM) network to model temporal gesture dynamics. Experimental results show that the proposed model achieves an overall recognition accuracy of approximately $95\%$. The trained model is further converted to TensorFlow Lite for real-time inference. This demonstrates the feasibility of the system for practical sign language translation applications.


\end{abstract}
\section{Introduction}
The design and development of effective tools for translating sign language are increasingly necessary to promote inclusive communication between deaf or hard-of-hearing individuals and the broader community. Sign languages are complex visual-spatial languages that rely on hand movements, facial expressions, and body posture, which make real-time interpretation challenging for people who do not understand them \cite{cooper2011sign}. As a result, communication barriers often limit access to education, healthcare, public services, and employment opportunities for the deaf community. Recent advances in artificial intelligence, computer vision, and sensor technologies provide new opportunities to automatically detect and interpret hand gestures for sign language translation.

Two primary approaches have dominated the development of sign language translation systems: vision-based approaches and glove-based approaches. Vision-based approaches employ cameras together with computer vision and machine learning algorithms to interpret hand gestures directly from visual data \cite{aloysius2020understanding,sharma2020vision}. By analyzing video frames or image sequences, these systems detect and track hand movements, extract spatial and temporal features, and map the observed gestures to corresponding words or sentences. With the rapid progress of deep learning, particularly convolutional neural networks \cite{9406809} 
and transformer-based models \cite{LIU2024106091}, vision-based methods have achieved significant improvements in gesture recognition accuracy and robustness \cite{adaloglou2021comprehensive,wadhawan2020deep}. However, these approaches often face practical limitations, including sensitivity to lighting variations, background clutter, occlusions, and changes in camera viewpoint. Advanced sensors, such as RGB-D cameras and LiDAR \cite{nguyen2024socially,nguyen2025real}, can partially mitigate these issues. Nevertheless, continuous tracking of fine-grained hand movements remains computationally demanding and may require carefully controlled recording conditions. 

These challenges have motivated the exploration of glove-based approaches for sign language recognition. Unlike vision-based systems, glove-based solutions rely on wearable sensors to directly capture hand motion and finger articulation, enabling more stable and precise measurements for gesture interpretation \cite{saeed2022systematic,tubaiz2015glove}. Such systems typically incorporate multiple sensing components such as flex sensors and inertial measurement units (IMUs) to measure finger bending, hand orientation, and motion dynamics \cite{wang2024wearable}. Early work demonstrated that instrumented gloves could capture detailed hand articulation and convert gestures into textual or spoken outputs to provide more reliable recognition compared with camera-based systems under challenging environmental conditions \cite{phi2015glove,tubaiz2015glove}. Later studies have further improved sensing accuracy by introducing advanced sensing materials, integrating multiple sensor modalities, and applying machine learning algorithms for dynamic gesture classification. For example, the use of soft fiber-optic sensors combined with machine learning achieved high recognition accuracy by capturing fine-grained finger motion patterns \cite{zhu2023machine}. Integration flex and IMU sensors with deep learning techniques allowed to recognize complex sign language vocabularies in real time \cite{feng2023design}. Despite these advantages, glove-based systems may introduce usability challenges, including the need for wearable hardware, calibration requirements, and potential discomfort during prolonged use. Nevertheless, they remain an important research direction for achieving accurate and robust sign language translation, particularly in environments where vision-based approaches are less reliable \cite{joksimoski2022technological,saeed2022systematic}.

In this work, we propose a multimodal system that combines glove-based sensing and vision-based analysis for Vietnamese sign language (VSL) recognition. The system uses a smart glove equipped with flex sensors and an IMU to capture finger bending, hand shape, and three-dimensional hand motion. Sensor data are collected by an ESP32-C6 microcontroller and transmitted to a computer, where they are fused with facial expression information captured by the computer’s camera. A deep learning model processes these multimodal inputs to infer sign language gestures. By integrating wearable motion sensing with visual facial cues, the proposed system captures both manual and non-manual components of sign language to improve recognition accuracy and robustness.

\section{System architecture}
Sign languages across different countries share common characteristics, as they rely on a combination of hand movements, gestures, and facial expressions to convey meaning. Accordingly, recognizing sign language requires capturing both the structural sequencing of gestures and the temporal dynamics of hand motion, as well as non-manual cues such as facial expressions.

\subsection{System overview}

\begin{figure}[ht]
\centering
\includegraphics[width=0.7\textwidth]{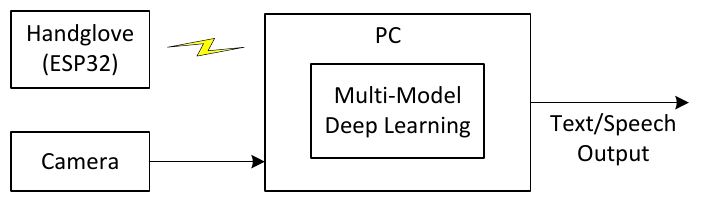}
\caption{System overview}
\label{fig:system}
\end{figure}

The overall architecture of the proposed sign language recognition system is illustrated in Fig.~\ref{fig:system}. The system combines a wearable sensing glove and a vision-based module to capture both manual and non-manual components of Vietnamese Sign Language (VSL).

The sensing glove integrates flex sensors and an IMU to measure finger articulation and hand motion. Sensor data are collected by an embedded microcontroller and transmitted via Wi-Fi to a host computer. In parallel, a camera connected to the computer captures facial expressions that provide additional contextual information in sign language communication.

On the computer, a multimodal deep learning model processes both the sensor data from the glove and the visual information from the camera. By fusing these inputs, the system interprets the performed gesture and translates it into corresponding text or speech output.

\subsection{Hand glove hardware design}

\begin{figure}[ht]
\centering
\includegraphics[width=0.7\textwidth]{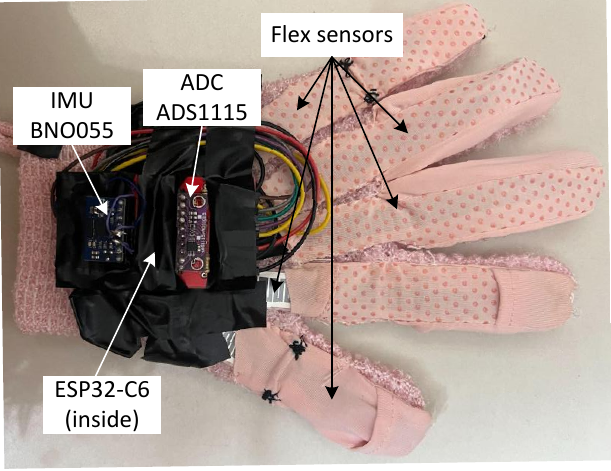}
\caption{Hardware design of the hand glove}
\label{fig:glove}
\end{figure}

The hardware design of the proposed sensing glove is shown in Fig.~\ref{fig:glove}. The glove is designed to capture detailed finger movements and hand orientation using multiple embedded sensors connected to an ESP32-C6 microcontroller.

\begin{itemize}

\item \textbf{Flex sensors:} Multiple SF15 flex sensors are positioned along the fingers of the glove to measure finger bending. Each sensor changes its resistance according to the degree of flexion. The resistance variation is converted into voltage signals using a voltage divider circuit and digitized by an ADS1115 analog-to-digital converter (ADC). The digitized measurements provide detailed information about finger articulation and hand shape for sign language recognition.

\item \textbf{IMU sensor:} An IMU BNO055 is mounted on the back of the hand to track hand orientation and motion. The BNO055 integrates an accelerometer, gyroscope, and magnetometer to estimate hand rotation and movement in three-dimensional space, which allow the system to capture dynamic gestures and spatial hand movements.

\item \textbf{Micro controller:} Sensor readings from the flex sensors and IMU are collected by an ESP32-C6 microcontroller located inside the glove enclosure. The ESP32-C6 performs sensor interfacing and data acquisition, and communicates with the ADS1115 and BNO055 via the I\textsuperscript{2}C interface. The collected data are then transmitted wirelessly to the host computer for further processing.

\end{itemize}

\begin{figure}[ht]
\centering
\includegraphics[width=\textwidth]{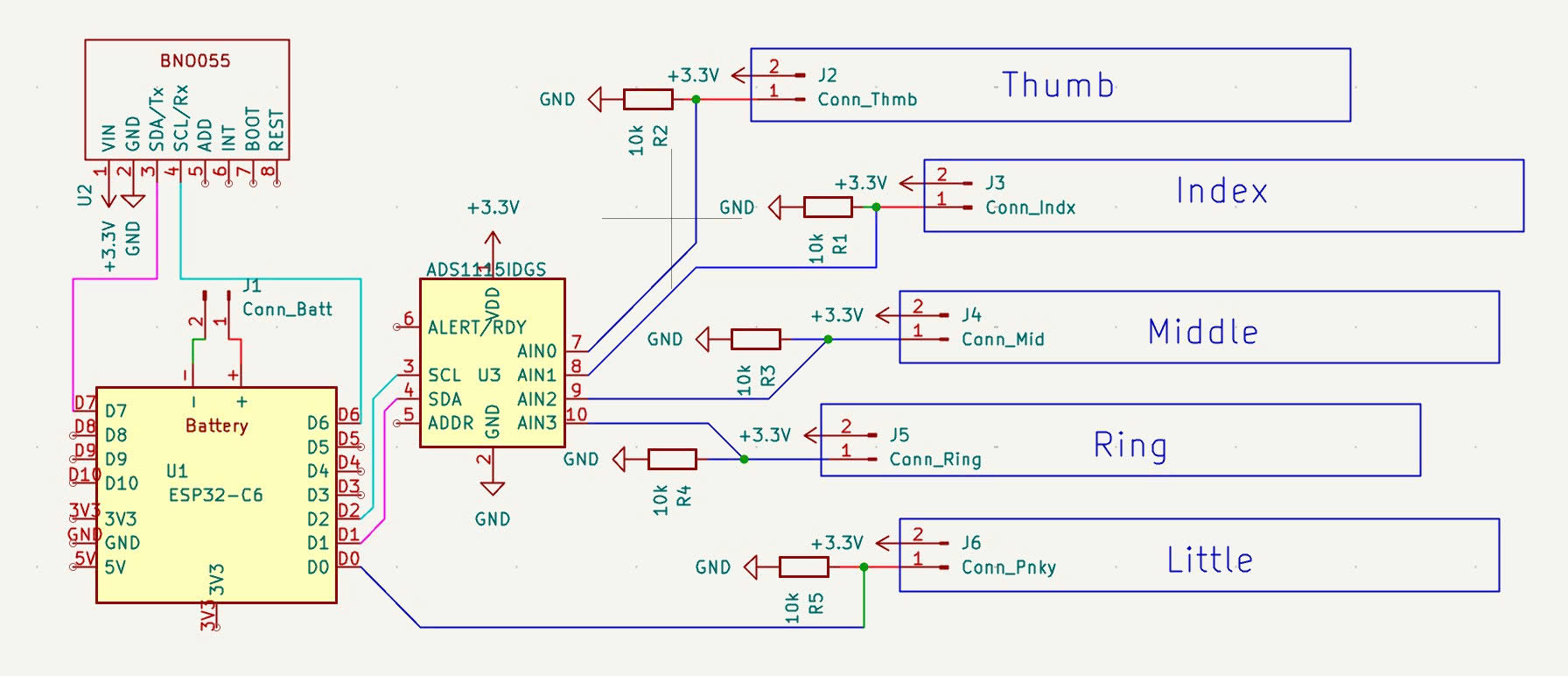}
\caption{Schematic circuit of the hand glove}
\label{fig:schematic}
\end{figure}

Figure~\ref{fig:schematic} shows the schematic circuit of the sensing glove. Each flex sensor is connected in a voltage divider configuration with a fixed resistor to convert resistance changes into measurable voltage signals. These analog signals are connected to the input channels of the ADS1115 ADC, which converts them into digital values. The ADS1115 communicates with the ESP32-C6 via the I\textsuperscript{2}C interface. The BNO055 IMU is also connected through the same I\textsuperscript{2}C bus to provide hand orientation and motion data. Power is supplied through a battery module that powers the ESP32-C6 and the connected sensors.

\section{Multimodal deep learning model}

\begin{figure}[ht]
\centering
\includegraphics[width=\linewidth]{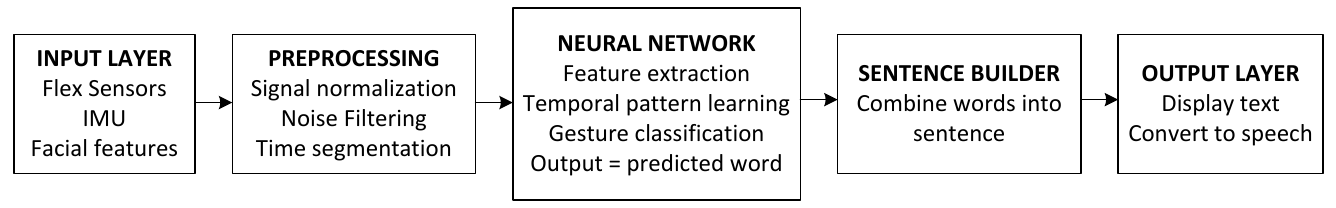}
\caption{System pipeline for gesture recognition, including input acquisition, preprocessing, neural network inference, sentence construction, and output generation.}
\label{fig:gesture_pipeline}
\end{figure}

The proposed system adopts a multimodal deep learning architecture that integrates hand motion signals and facial cues for gesture recognition. As illustrated in Fig.~\ref{fig:gesture_pipeline}, the pipeline consists of five main stages: input acquisition, signal preprocessing, neural network inference, sentence construction, and output generation.

\subsection{Input layer and data preprocessing}

The input layer collects multimodal data from three sensing sources: flex sensors (SF15), an IMU (BNO055), and camera-based facial analysis. These signals provide complementary information about hand configuration, motion dynamics, and contextual facial expressions. Raw sensor data are streamed continuously and timestamped during acquisition. To ensure temporal consistency across samples, the signals are uniformly resampled by retaining measurements separated by $80\,\text{ms}$ intervals. Each gesture instance consists of $25$ temporal frames, resulting in an approximate sample duration of

\[
25 \times 80\,\text{ms} = 2000\,\text{ms} \approx 2\,\text{s}.
\]
During preprocessing, the signals are normalized and filtered to reduce noise, and the continuous data stream is segmented into fixed-length gesture samples suitable for model training.

\noindent
For facial analysis, the system uses MediaPipe\cite{lugaresi2019mediapipe} to detect faces and extract landmarks, including the eyebrows, eyes, and mouth. The distances between these landmarks are computed and aggregated into a feature set. The values are then averaged within a defined range. Finally, the features are encoded into discrete values from 0 to 6, representing seven distinct expressions corresponding to grammatical components.

\subsection{Training data representation}

Let $T$ denote the number of temporal frames and $d$ denote the feature dimension. Each gesture instance is represented as a multivariate time-series $X \in \mathbb{R}^{T \times d}$, where $T = 25$ and the feature dimension is $d = 14$. The feature vector consists of the following measurements:

\begin{itemize}
\item {Flex sensors:} $flex_1, flex_2, flex_3, flex_4, flex_5, flex_6, flex_7, flex_8, flex_9, flex_{10}$
\item {IMU orientation signals:} $imu_x, imu_y, imu_z$
\item {Facial expression indicator:} $exp$
\end{itemize}

Thus, each gesture sample is represented as a sequence of $25$ frames with $14$ features per frame.

\subsection{Neural network architecture}
\begin{figure}[h]
\centering
\includegraphics[width=\textwidth]{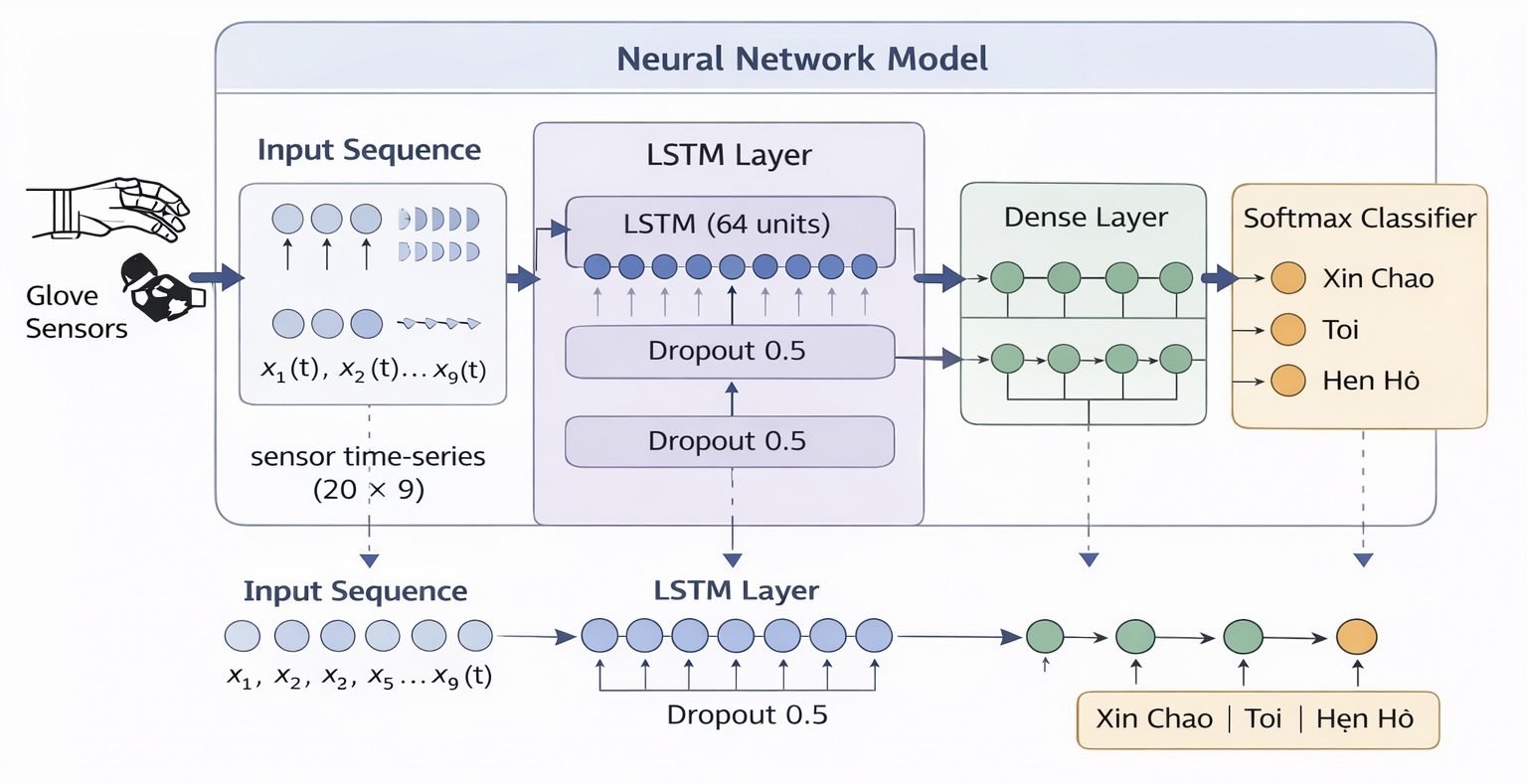}
\caption{Neural network inference pipeline for sign language recognition.}
\label{fig:pipeline}
\end{figure}

To model temporal dependencies within the input gesture sequence, the system employs a Long Short-Term Memory (LSTM) neural network, as shown in Fig. \ref{fig:pipeline}. LSTM networks are well suited for sequential data because they can learn temporal patterns and motion dynamics across time steps.

The first LSTM layer contains 64 hidden units and returns the full output sequence. This allows the network to preserve temporal information across all time steps. A dropout layer with a dropout rate of 0.5 is applied to reduce overfitting during training.

The output sequence is then processed by a second LSTM layer with 64 hidden units, which encodes the temporal sequence into a compact feature representation. Another dropout layer with a rate of 0.5 is applied for additional regularization.

The resulting representation is passed to a fully connected dense layer with 64 neurons and ReLU activation to learn higher-level feature interactions. Finally, a softmax output layer produces a probability distribution over the gesture classes.

The network is trained using the Adam optimizer and the categorical cross-entropy loss function, which measures the discrepancy between the predicted class probabilities and the ground-truth labels. The loss function is defined as

\begin{equation}
L = - \sum_{i=1}^{C} y_i \log(\hat{y}_i),
\end{equation}
where \(C\) denotes the number of gesture classes, \(y_i\) represents the ground-truth label in one-hot encoded form, and \(\hat{y}_i\) is the predicted probability produced by the softmax layer. Training is performed for up to 50 epochs with a batch size of 128. A validation split of 20\% is used to monitor training performance, and early stopping with a patience of five epochs is applied to prevent overfitting.

\section{Results}
A number of experiments were conducted to evaluate the performance of the proposed approach, as described below.

\subsection{Dataset preparation}

The dataset used in the experiments was collected using the proposed sign language glove system. Each gesture recording consists of synchronized readings from multiple onboard sensors, including five flex sensors embedded along the fingers and an IMU providing motion data along the $x$, $y$, and $z$ axes. An additional signal labeled \textit{face} was also recorded to capture supplementary gesture-related information.

Three Vietnamese sign language gestures were included in the dataset: ```Tôi'', ``Xin chào'', and ``Hẹn hò''. Multiple recordings were collected for each gesture, resulting in a total of 10 gesture sequences. Each sequence contains 25 time steps, representing the temporal evolution of the hand gesture during its execution.

At each time step, a 9-dimensional feature vector is recorded:
\begin{equation}
[f_1, f_2, f_3, f_4, f_5, x, y, z, face]
\end{equation}
where $f_1$-$f_5$ correspond to the five flex sensor readings and $x$, $y$, and $z$ represent the IMU motion measurements. Consequently, each gesture sample forms a $25 \times 9$ time-series matrix. This structure enables the learning model to capture both finger articulation patterns and hand motion dynamics during gesture execution. Table~\ref{tab:dataset_example} shows a short excerpt of the recorded sensor data for one gesture sequence.

\begin{table}[h]
\centering
\caption{Example of recorded sensor data for a gesture sequence}
\label{tab:dataset_example}
\begin{tabular}{c c c c c c c c c c}
\hline
$t$ & $f_1$ & $f_2$ & $f_3$ & $f_4$ & $f_5$ & $x$ & $y$ & $z$ & face \\
\hline
1 & 520 & 498 & 470 & 455 & 430 & -0.12 & 0.05 & 0.98 & 1 \\
2 & 523 & 500 & 472 & 457 & 431 & -0.10 & 0.06 & 0.97 & 1 \\
3 & 528 & 503 & 475 & 460 & 435 & -0.09 & 0.07 & 0.96 & 1 \\
4 & 531 & 505 & 478 & 463 & 437 & -0.08 & 0.08 & 0.95 & 1 \\
\hline
\end{tabular}
\end{table}

Overall, the dataset contains 250 time steps of sensor measurements across all recordings. The temporal nature of the data makes it suitable for sequence learning models such as LSTM networks.

\begin{figure}[ht]
\centering
\begin{subfigure}{0.5\textwidth}
    \centering
    \includegraphics[width=\linewidth]{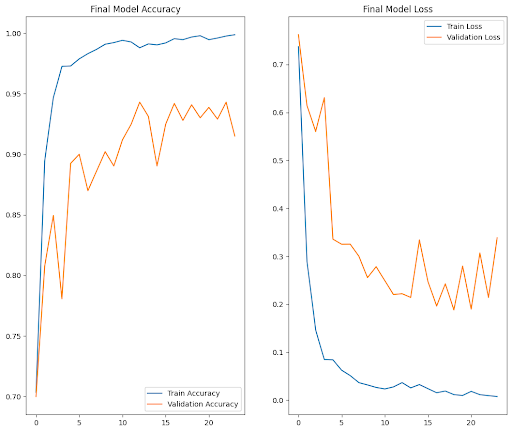}
    \caption{Training and validation metrics}
    \label{fig:training_metrics}
\end{subfigure}
\hfill
\begin{subfigure}{0.45\textwidth}
    \centering
    \includegraphics[width=\linewidth]{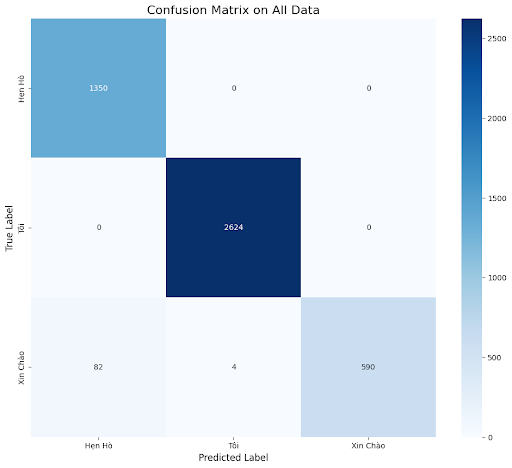}
    \caption{Confusion matrix on trained data}
    \label{fig:confusion_matrix}
\end{subfigure}
\caption{Recognition result of the proposed model.}
\label{fig:training_results}
\end{figure}

\subsection{Training results}

The proposed LSTM-based gesture recognition model was trained using the collected sign language glove dataset. During training, both accuracy and loss were monitored for the training and validation sets to evaluate the learning behavior and generalization performance of the network.

Fig.~\ref{fig:training_metrics} shows the training and validation accuracy and loss curves over the training epochs. The training accuracy rapidly increases during the first few epochs and gradually converges to nearly $100\%$. The validation accuracy stabilizes around $93\%$-$95\%$, indicating that the model is able to generalize well to unseen samples. Meanwhile, the training loss decreases steadily toward zero, while the validation loss stabilizes at a low value, suggesting that the model successfully learns the temporal patterns of the gesture sequences without severe overfitting.

The confusion matrix shown in Fig.~\ref{fig:confusion_matrix} illustrates the classification performance across the three gesture classes: \textit{Hẹn Hò}, \textit{Tôi}, and \textit{Xin Chào}. The model achieves perfect classification for the gestures \textit{Hẹn Hò} and \textit{Tôi}, with all samples correctly predicted. For the \textit{Xin Chào} gesture, the majority of samples are correctly classified, while a small number are misclassified as \textit{Hẹn Hò} and \textit{Tôi}.

Overall, the model achieves the accuracy of approximately $95\%$. These results demonstrate that the proposed model can effectively capture both finger articulation patterns from the flex sensors and motion dynamics from the IMU data, enabling reliable recognition of VSL gestures.

\begin{figure}[h]
\centering
\includegraphics[width=0.6\textwidth]{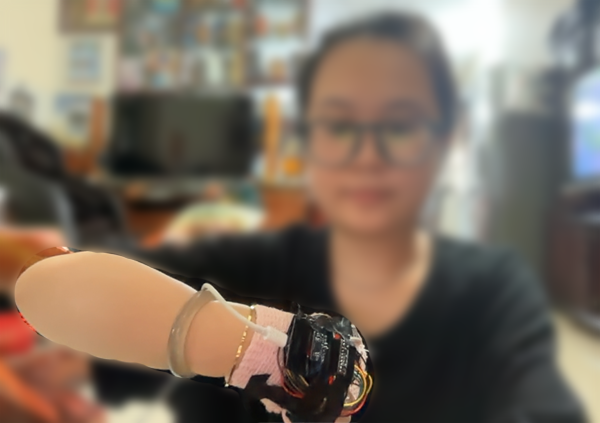}
\caption{Real-time recognition using the proposed sign language glove.}
\label{fig:experiment}
\end{figure}

\subsection{Real-time deployment on hardware}

To evaluate the practical applicability of the proposed system, the trained gesture recognition model was deployed on a real hardware setup using the sign language glove. The trained model was converted to TensorFlow Lite format and executed on a laptop connected to the glove system for real-time inference. 

During testing, the glove collects sensor data from the flex sensors and inertial measurement unit while the user performs a gesture, as shown in Fig.\ref{fig:experiment}. In this example, the user performs the gesture corresponding to ``Tôi''. The system successfully records the sensor sequence, loads the trained model, and performs inference in real time. The model correctly predicts the gesture as ``Tôi'' with a confidence score of $95.34\%$. 

In addition, the glove was tested on five participants with different hand sizes. Experiments were conducted under varying lighting conditions and distances. Optimal facial recognition performance was achieved when the user was positioned approximately one arm’s length from the camera. Users also reported that the system was comfortable to use during testing.

The results demonstrate that the proposed gesture recognition system can operate in a real-world environment. The deployment confirms that the trained LSTM model is lightweight enough for real-time execution while maintaining high prediction accuracy.

\section{Conclusion}

This paper presented a multimodal smart glove system for sign language recognition that integrates wearable sensing and deep learning. The proposed glove uses flex sensors and an IMU to capture finger articulation and hand motion, while facial cues are obtained from a camera. A LSTM network processes the multimodal data to recognize gesture sequences. Experimental results show that the system achieves approximately $95\%$ recognition accuracy. The trained model was converted to TensorFlow Lite and successfully deployed for real-time inference. Future work will focus on expanding the gesture vocabulary, collecting larger datasets, and improving system portability.

\bibliographystyle{splncs04}
\bibliography{ref}

@article{LIU2024106091,
title = {Are transformer-based models more robust than CNN-based models?},
journal = {Neural Networks},
volume = {172},
pages = {106091},
year = {2024},
issn = {0893-6080},
author = {Zhendong Liu and Shuwei Qian and Changhong Xia and Chongjun Wang},
}

@ARTICLE{9406809,
  author={Zhu, Qiuchen and Dinh, Tran Hiep and Phung, Manh Duong and Ha, Quang Phuc},
  journal={IEEE Access}, 
  title={Hierarchical Convolutional Neural Network With Feature Preservation and Autotuned Thresholding for Crack Detection}, 
  year={2021},
  volume={9},
  number={},
  pages={60201-60214},
  }

@article{nguyen2025real,
  title={Real-time recognition of human interactions from a single RGB-D camera for socially-aware robot navigation},
  author={Nguyen, Thanh Long and Nguyen, Duc Phu and Nu, Thanh Thao Ton and Le, Quan and Tran, Thuan Hoang and Phung, Manh Duong},
  journal={Intelligent Service Robotics},
  volume={18},
  number={6},
  pages={1369--1380},
  year={2025},
  publisher={Springer}
}

@inproceedings{nguyen2024socially,
  title={Socially aware motion planning for service robots using lidar and RGB-d camera},
  author={Nguyen, Duc Phu and Nguyen, Thanh Long and Tu, Minh Dang and Quach, Cong Hoang and Truong, Xuan Tung and Phung, Manh Duong},
  booktitle={2024 International Conference on Control, Robotics and Informatics (ICCRI)},
  pages={109--114},
  year={2024},
  organization={IEEE}
}

@inproceedings{lugaresi2019mediapipe,
  title={Mediapipe: A framework for perceiving and processing reality},
  author={Lugaresi, Camillo and Tang, Jiuqiang and Nash, Hadon and McClanahan, Chris and Uboweja, Esha and Hays, Michael and Zhang, Fan and Chang, Chuo-Ling and Yong, Ming and Lee, Juhyun and others},
  booktitle={Third workshop on computer vision for AR/VR at IEEE computer vision and pattern recognition (CVPR)},
  volume={2019},
  pages={2},
  year={2019},
  organization={Long Beach, CA}
}

@incollection{cooper2011sign,
  title={Sign language recognition},
  author={Cooper, Helen and Holt, Brian and Bowden, Richard},
  booktitle={Visual analysis of humans: Looking at People},
  pages={539--562},
  year={2011},
  publisher={Springer}
}

@inproceedings{sharma2020vision,
  title={Vision-based sign language recognition system: A Comprehensive Review},
  author={Sharma, Sakshi and Singh, Sukhwinder},
  booktitle={2020 international conference on inventive computation technologies (ICICT)},
  pages={140--144},
  year={2020},
  organization={IEEE}
}

@article{aloysius2020understanding,
  title={Understanding vision-based continuous sign language recognition},
  author={Aloysius, Neena and Geetha, Madathilkulangara},
  journal={Multimedia Tools and Applications},
  volume={79},
  number={31},
  pages={22177--22209},
  year={2020},
  publisher={Springer}
}

@article{tubaiz2015glove,
  title={Glove-based continuous Arabic sign language recognition in user-dependent mode},
  author={Tubaiz, Noor and Shanableh, Tamer and Assaleh, Khaled},
  journal={IEEE Transactions on Human-Machine Systems},
  volume={45},
  number={4},
  pages={526--533},
  year={2015},
  publisher={IEEE}
}

@article{saeed2022systematic,
  title={A systematic review on systems-based sensory gloves for sign language pattern recognition: An update from 2017 to 2022},
  author={Saeed, Zinah Raad and Zainol, Zurinahni Binti and Zaidan, BB and Alamoodi, Abdullah Hussein},
  journal={IEEE Access},
  volume={10},
  pages={123358--123377},
  year={2022},
  publisher={IEEE}
}

@article{wadhawan2020deep,
  title={Deep learning-based sign language recognition system for static signs},
  author={Wadhawan, Ankita and Kumar, Parteek},
  journal={Neural computing and applications},
  volume={32},
  number={12},
  pages={7957--7968},
  year={2020},
  publisher={Springer}
}

@article{adaloglou2021comprehensive,
  title={A comprehensive study on deep learning-based methods for sign language recognition},
  author={Adaloglou, Nikolas and Chatzis, Theocharis and Papastratis, Ilias and Stergioulas, Andreas and Papadopoulos, Georgios Th and Zacharopoulou, Vassia and Xydopoulos, George J and Atzakas, Klimnis and Papazachariou, Dimitris and Daras, Petros},
  journal={IEEE transactions on multimedia},
  volume={24},
  pages={1750--1762},
  year={2021},
  publisher={IEEE}
}

@article{wang2024wearable,
  title={Wearable electronic glove and multilayer para-LSTM-CNN-based method for sign language recognition},
  author={Wang, Dapeng and Wang, Mingyuan and Zhang, Ziqi and Liu, Teng and Meng, Chuizhou and Guo, Shijie},
  journal={IEEE Internet of Things Journal},
  volume={11},
  number={24},
  pages={40787--40799},
  year={2024},
  publisher={IEEE}
}

@inproceedings{phi2015glove,
  title={A glove-based gesture recognition system for Vietnamese sign language},
  author={Phi, Lam T and Nguyen, Hung D and Bui, TT Quyen and Vu, Thang T},
  booktitle={2015 15th International Conference on Control, Automation and Systems (ICCAS)},
  pages={1555--1559},
  year={2015},
  organization={IEEE}
}

@article{zhu2023machine,
  title={Machine-learning-assisted soft fiber optic glove system for sign language recognition},
  author={Zhu, Renjie and Fan, Dongliang and Lin, Jiang and Feng, Huijuan and Wang, Hongqiang and Dai, Jian S},
  journal={IEEE Robotics and Automation Letters},
  volume={9},
  number={2},
  pages={1540--1547},
  year={2023},
  publisher={IEEE}
}

@article{feng2023design,
  title={Design and implementation of gesture recognition system based on flex sensors},
  author={Feng, Deli and Zhou, Cheng and Huang, Jipeng and Luo, Gangyin and Wu, Xin},
  journal={IEEE Sensors Journal},
  volume={23},
  number={24},
  pages={31389--31398},
  year={2023},
  publisher={IEEE}
}

@article{joksimoski2022technological,
  title={Technological solutions for sign language recognition: a scoping review of research trends, challenges, and opportunities},
  author={Joksimoski, Boban and Zdravevski, Eftim and Lameski, Petre and Pires, Ivan Miguel and Melero, Francisco Jos{\'e} and Martinez, Tom{\'a}s Puebla and Garcia, Nuno M and Mihajlov, Martin and Chorbev, Ivan and Trajkovik, Vladimir},
  journal={IEEE Access},
  volume={10},
  pages={40979--40998},
  year={2022},
  publisher={IEEE}
}

\end{document}